\documentclass[letter]{aa}  
\usepackage{graphicx}
\usepackage{txfonts}
\usepackage{natbib}
\usepackage[citecolor=blue,colorlinks=true,linkcolor=blue,urlcolor=blue]{hyperref}
\graphicspath{{./}}
\usepackage[export]{adjustbox}
\usepackage{tablefootnote}

\begin{document} 

\title{Fe\,K$\alpha$ line from the Broad Line Region of PDS456 with XRISM/Resolve}

\author{Alfredo Luminari
\inst{1},\inst{2}\thanks{\email{alfredo.luminari@inaf.it}}
\and Fabrizio Nicastro \inst{1}
\and Stefano Bianchi\inst{3}
\and Elias Kammoun\inst{4}}

\institute{
INAF – Osservatorio Astronomico di Roma, Via Frascati 33, 00078 Monteporzio, Italy
\and
INAF – Istituto di Astrofisica e Planetologia Spaziali, Via del Fosso del Caveliere 100, 00133 Roma, Italy
\and
Dipartimento di Matematica e Fisica, Universit\'{a} degli Studi Roma Tre, Via della Vasca Navale 84, I-00146, Roma, Italy
\and 
Cahill Center for Astronomy \& Astrophysics, California Institute of Technology, 1216 East California Boulevard, Pasadena, CA 91125, USA
}
\date{Received 09/06/2026 / Accepted 11/07/2026}

\abstract{The high-luminosity, high-mass and likely super-Eddington quasar PDS 456 is known for its extremely dynamic environment. A wealth of observational features from the millimetric up to the X-ray band shows energetic outflows at all scales, from accretion disc up to galaxy scales. Broad Emission Lines in the optical and UV show significant widths and line-of-sight velocities. Moreover, NIR interferometry revealed a partially outflowing Broad Line Region. 

Thanks to the unprecedented energy resolution of the X-ray microcalorimeter Resolve onboard XRISM, we detect a neutral Fe K$\alpha$ line, the main tracer of cold matter around AGNs. The line blueshift ($v_{\rm out}=2700$ km s$^{-1}$) and width ($\sigma \leq 700$ km s$^{-1}$) are on the lower bound of the range of values of the optical-UV emission lines, possibly suggesting a stratified medium. The derived column density, $N_{\rm H} \approx 10^{22}$ cm$^{-2}$, is significantly lower than what expected for Broad Line Regions, again in agreement with a composite emission.
The very small Equivalent Width (9 eV) marks one of the smallest measurements in the literature and probes the high-luminosity end of the so-called Iwasawa-Taniguchi (or X-ray Baldwin) effect.}

\authorrunning{A. Luminari et al.}
\titlerunning{Fe\,K$\alpha$ line from the Broad Line Region of PDS456 with XRISM/Resolve}

\maketitle

\section{Introduction \label{sec:intro}}
The bright quasar PDS\,456 ($z_{\rm PDS}$=0.185, \citealp{2019A&A...628A.118B}) stands out as the most luminous Active Galactic Nucleus (AGN) in the nearby ($z \leq 2$) Universe, with a bolometric luminosity $L_{\rm bol} \approx 10^{47}\,\rm erg\ s^{-1}$ \citep{1997ApJ...488L..19T,nrg15}. Its accretion rate is likely higher than the Eddington rate (\citealp{2024A&A...690A..76G,2025Natur.641.1132X}, hereafter X25) and outflowing gas is detected at all spatial scales and wavelengths. \cite{2019A&A...628A.118B} detected clumpy outflows in ALMA observations of CO(3-2) line extending at galaxy scales. The gas depletion time is $\sim 8 \,\rm Myr$, indicating that the outflows are able to sweep the galactic gas reservoir, potentially halting the star formation. Likewise, \cite{2024A&A...686A.250T} detected powerful winds in the optical band with MUSE in [\ion{O}{iii}] and H$\alpha$ transitions, extending even beyond the host galaxy. 

In the X-ray band, PDS\,456 hosts the most clear-cut example of a \ion{Fe}{xxv-xxvi} P-Cygni profile due to a massive, highly ionised wind originating at accretion disc scales \citep{nrg15, lpt18}. Such powerful outflow has been recently detected at high resolution with the microcalorimeter Resolve \citep{tashiro20} onboard XRISM (X25). The huge improvement with respect to previous, CCD-based observations finally resolved the broad absorption through in a number of narrow layers with line-of-sight (LOS) velocities $v_{\rm out}=0.22 - 0.33$ c, each one with column density $N_{\rm H}\approx 10^{23}\,\rm cm^{-2}$ and line broadening $\sigma_{\rm turb} \approx 2000\,\rm km\ s^{-1}$. The global mass outflow rate is of the order of $100\,\rm M_{\rm \odot}\ yr^{-1}$, resulting in a kinetic power $\dot{E}_{\rm kin} \approx 10^{47}\rm erg\, s^{-1}$, of the same order of $L_{\rm bol}$ \citep{nrg15}. Such huge energy output is well above the $\dot{E}_{\rm kin} \geq 0.5\%-5\%\, L_{\rm bol}$ theoretical threshold for the outflows to efficiently impact the host galaxy \citep{he10,Faucher12,King15}. However, the galactic-scale outflow counterparts detected at optical to millimetric wavelengths are way less energetic than expected, being more consistent with a momentum-conserving propagation rather than with an energy-conserving one, possibly signalling different propagation scenarios at such high luminosity regimes (see \citealp{2019A&A...628A.118B,2024A&A...686A.250T} for further details). Outflows have also been detected with the JWST \citep{2024ApJ...976..240S}, mainly from dust and molecular lines, as well as Pa$\alpha$, [\ion{O}{iii}], [\ion{Ne}{iii}] and  [\ion{Ne}{vi}].

Broad Emission Lines (BELs) in PDS 456 show blueshifted profiles with a quite high broadening, with a full width at half maximum (FWHM) ranging between $3500\,\rm km\,s^{-1}$ for the Hydrogen Balmer series \citep{1999MNRAS.303L..23S} and  $15000\,\rm km\,s^{-1}$ for \ion{C}{iv} \citep{2005MNRAS.360L..25O}. Near-infrared interferometry carried out with VLT/GRAVITY spatially resolved the Pa$\alpha$ line from the Broad Line Region (BLR), finding evidence for a partially bound gas, with LOS outflow velocities up to $2000\,\rm km\,s^{-1}$. Therefore, all the spectral features, from the NIR up to the UV, lead to the picture of an outflowing BLR. While this is not at all unprecedented \citep[see discussion in][and references therein]{2024A&A...690A..76G} , such extreme velocities may suggest that PDS\,456 is probably probing the BLR behaviour at extreme quasar luminosities.

In this Letter we report the first detection of a blueshifted fluorescent Fe\,K$\alpha$ line in PDS 456, observed in the Resolve spectrum first reported in X25. This adds a further intriguing piece in the puzzle of the dynamic structure of this quasar and represents a valuable probe of the dense matter at the high-luminosity end of the AGN population.

\section{Data reduction} \label{sec:Data_reduction}
XRISM observed PDS\,456 during the Performance Verification Phase, from 11 to 17 March 2024. Data were retrieved from the mission archive\footnote{\url{https://darts.isas.jaxa.jp/}} and processed with the latest available pipeline, provided within the Heasoft v6.36\footnote{\url{https://heasarc.gsfc.nasa.gov/docs/software/lheasoft/}}, and calibration database (CALDB Resolve files v.20250915).
The Resolve data and the Non X-Ray Background (NXB) were reduced following standard prescriptions\footnote{See \url{https://heasarc.gsfc.nasa.gov/docs/xrism/analysis/}} and excluding the calibration pixel, no. 12, and pixel no. 27 due to known gain calibration issues. We extracted the spectrum with different thresholds for the geomagnetic cut-off rigidity (COR). Setting COR>8, the most conservative choice, results in 187\,ksec net exposure time, while $\rm COR>4$ yields 260 ksec. Since the two datasets are fully consistent within each other, we use the latter from now on (see Appendix \ref{appendix} for more details).

In this Letter we focus on the Fe\,K$\alpha$ feature. Therefore, we do not report on the complementary \textit{XRISM}/Xtend, \textit{XMM-Newton} and \textit{NuSTAR} data, since their CCD energy resolutions of hundreds of eV do not allow a meaningful detection of this feature which, as will be shown in the following, has an Equivalent Width (EW) of the order of 9 eV. In the following we use the \texttt{xspec} fitting package \citep{xspec}. Unless when stated otherwise, we bin the data to 5 eV resolution, the nominal energy resolution of Resolve, and we employ the Cash statistics \citep{cash76}. However, consistent results are obtained binning to lower (2 eV) or higher (10 or 15 eV) energy intervals.

\section{Spectral Analysis} \label{sec:spectra}
\subsection{Broadband inspection}
The X-ray spectrum of PDS\,456 is notoriously complex. The intrinsic continuum is well described by a powerlaw with photon index $\Gamma=2.0 - 2.3$ and no evidence for reflection (\citealp{nrg15}, X25). This continuum is reprocessed by the many intervening winds, both in the soft and in the hard X-ray bands. The massive, highly ionised and mildly relativistic disc winds are responsible for the bright Fe\,K emission, mostly due to H- and He-like ions \citep{lpt18}. Due to the high outflow velocities (between 0.2 and 0.3 c, see above), the emission profile is relativistically-broadened and extends from 6 to 8 keV. Blueward of that, the high-resolution Resolve spectrum shows several highly-blueshifted Fe\,K absorption lines strongly affecting the underlying continuum up to around 10 keV, after which the spectrum becomes background-dominated (see Fig. 3 of X25). 
In order to visually inspect the spectrum, we first fit the whole $2-10$\,keV energy band with the following phenomenological model:
\begin{equation}
{\tt Model = TBabs \times (powerlaw + Gauss_{wind}).}
\label{cont_model}
\end{equation}
The model consists of a power-law continuum and a broad Gaussian emission line to account for the wind emission. The source spectrum is absorbed by a \texttt{TBabs} \citep{TBabs} cold absorption component with a Galactic column of $N_{\rm H}=2 \times 10^{21}\,\rm cm^{-2}$ (as per \citealp{HI4PI}). Hereafter, the NXB spectrum is always included and fitted jointly with the source spectrum using the empirical model made available by the Science Team\footnote{\url{https://heasarc.gsfc.nasa.gov/docs/xrism/analysis/nxb/resolve_nxb_db.html}}, which is composed by a flat power law ($\Gamma=0.14$) with several narrow neutral emission lines. The best-fit result is shown in Fig \ref{fig:cortime}. The power-law component has $\Gamma=2.07 \pm 0.08$ and normalisation $(1.6 \pm 0.2) \times \,\rm 10^{-3}\,\rm ph\,cm^{-2}\, s^{-1}\,keV^{-1}$ at 1\,keV. ${\tt Gauss_{\rm wind}}$ has line energy $=4.9 \pm 0.2\rm\,keV$, a broadening $\sigma=1.5 \pm0.2\rm\,keV$, and $\rm EW=3.4 \pm 0.7\rm\,keV$. The 2-10 keV flux is 5.6$\cdot 10^{-12}$ erg s$^{-1}$ cm$^{-2}$, corresponding to 5.5$\cdot 10^{44}$ erg s$^{-1}$.

By inspecting the residuals, we note an excess emission at $E\approx 5.45\,\rm keV$ (see Fig. \ref{fig:spectrum}, middle panel), corresponding to 6.46 keV source-frame, slightly blueshifted with respect to the energy of the Fe K$\alpha$ line. Therefore, we focus on a narrow spectral interval to accurately characterise such excess. 

\subsection{Narrow-band fit}

In the following, we restrict the analysis to the $3-6\rm\,keV$ energy range. Such interval is sufficiently narrow so that the continuum can be reliably described with the simple model in Eq. \ref{cont_model} and, yet, wide enough to allow a solid determination of it (see \citealp{2010ApJS..187..581S} for the same approach with high-resolution \textit{Chandra} grating spectra). This approach is only to present a good analytical description of the continuum and to characterise the narrow line on top of it. We do not attempt to draw any physical conclusions associated with the continuum properties from this analysis.

To model the Fe\,K$\alpha$ line we add a \texttt{zbfeklor} component, a composition of seven Lorentzians obtained as the empirical fit to high-resolution laboratory measurements of the fluoresecent neutral Fe K$\alpha$ line \citep{1997PhRvA..56.4554H}. Such component represents the standard in high-resolution spectroscopy and it is widely employed for high signal to noise Resolve data (see e.g. \citealp{2025PASJ...77.1210Y,2026arXiv260222476X,2026A&A...710A.274B}). The only free parameters are the normalisation, the overall velocity broadening $\sigma$ and the blue/red-shift $z$ of the line energy with respect to the laboratory value. We fit leaving all the parameters free to vary, obtaining a fit statistics of 601.6 for 592 degrees of freedom. Note that replacing \texttt{zbfeklor} with a Gaussian component would lead to fully consistent results but with a $\Delta$C-stat increase of 2.

\begin{figure}
\centering
\includegraphics[width=\columnwidth]{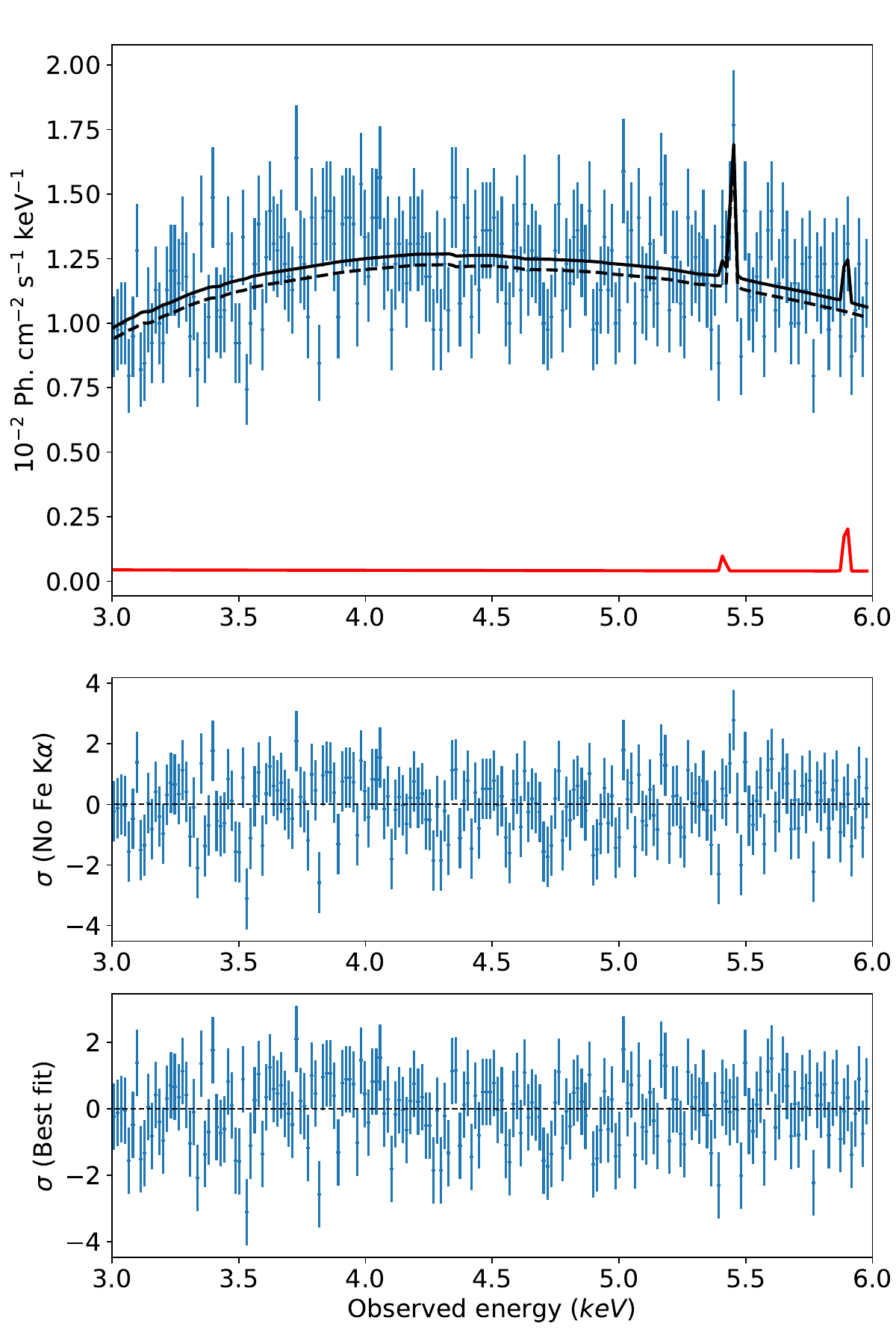}
\caption{Spectral fit of the $3-6$\,keV band. Top: data and best-fit model. Red line shows the NXB model, dashed black line the source model and black solid line the composition of the two. Centre and bottom panels: residuals (in units of $\sigma$) without and with the Fe\,K$\alpha$ line. Spectrum has been re-binned to 15\,eV for plotting purposes only.} 
\label{fig:spectrum}
\vspace{-2ex}
\end{figure}

\begin{figure}
\centering
\includegraphics[width=\columnwidth]{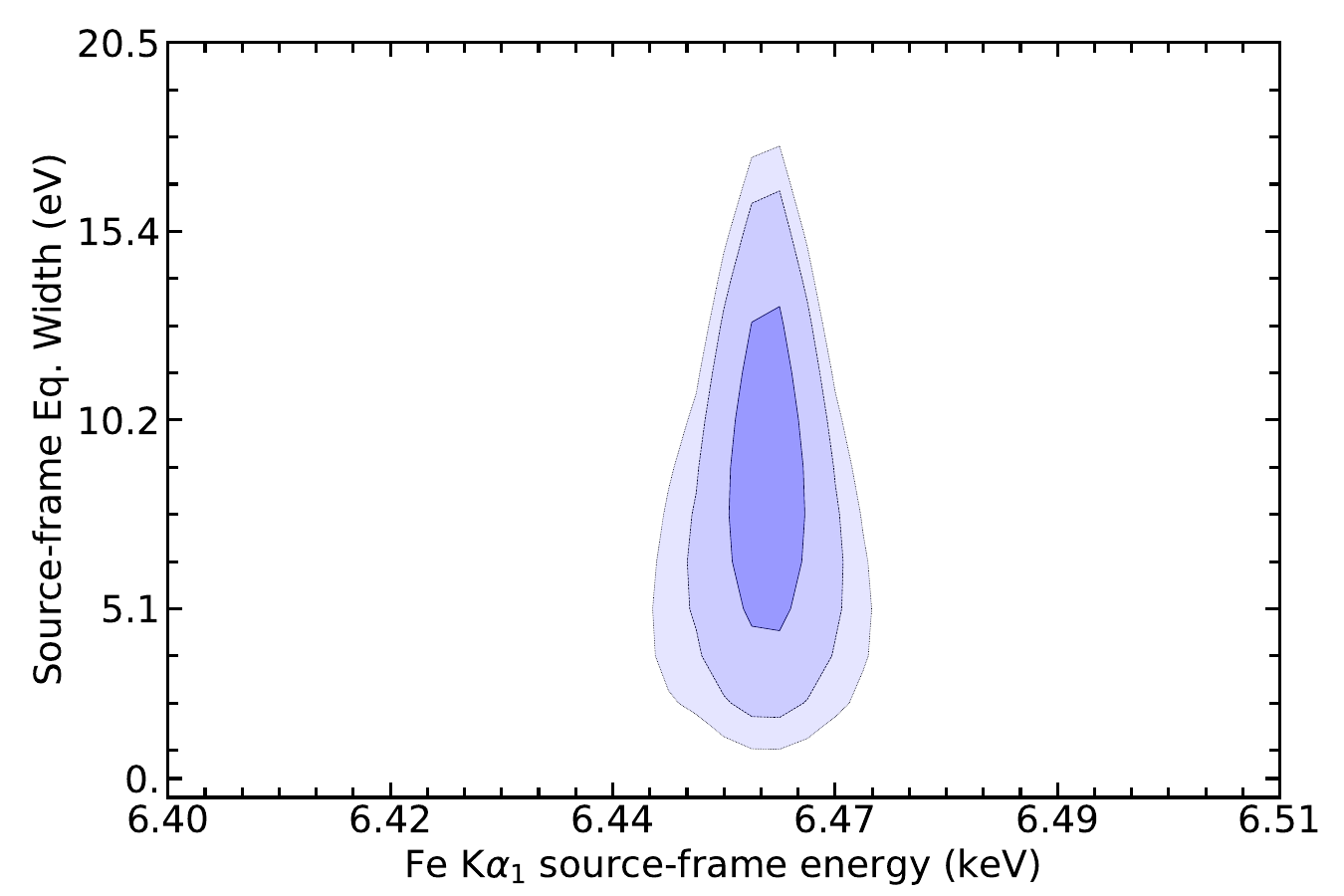}
\caption{Contour plot between the derived Fe\,K$\alpha_1$ line energy (x-axis) and the overall line Equivalent Width (y-axis), both in the PDS\,456 source frame. Confidence levels correspond to 1,2,3 $\sigma$ for two parameters of interest. } 
\label{fig:contour}
\end{figure}

Fig.\,\ref{fig:spectrum} shows the best fit and the residuals, both without and with the Fe\,K$\alpha$ line. The source model is plotted with a dashed black line, while the solid black line shows the total model including the NXB (in red). The nearest NXB line has an energy of 5.415\,keV, around 8 resolution elements from the peak of the observed line (5.452\,keV) and a factor $\approx 20$ lower strength and, therefore, it is quite negligible.
Table \ref{tab:fit} reports the best fit values with the associated 1 $\sigma$ uncertainties. The statistical improvement upon the inclusion of the Fe\,K$\alpha$ is $\Delta C_{\rm stat}=11.23$. 
The line EW in the source-frame is $9^{+4}_{-3}$ eV, obtained by dividing the observed one by (1+z$_{PDS}$). Thanks to the unprecedented energy resolution of Resolve, the upper limit of the line broadening is $\sigma \leq 310\,(694)\,\rm km \,s^{-1}$ at 1$\sigma$ (90 \%) c.l., corresponding to an energy width $\leq$ 5.6 (12.6)\,eV. Surprisingly, the redshift is smaller than the systemic one of PDS\,456 ($z_{\rm PDS}=0.185$), implying $v_{\rm out}=2700 \pm 300$ km s$^{-1}$ (90\% c.l.). To provide a more physical picture, Fig. \ref{fig:contour} shows the contour plot (with $1,2,3 \sigma$ confidence levels) between the derived Fe\,K$\alpha_1$ line energy (in the PDS\,456 source frame), and the total feature EW. Note that we are ascribing the feature to neutral iron (\ion{Fe}{ii}). As discussed in \cite{2003A&A...410..359P,2026A&A...710A.274B}, mildly ionised Fe would have lower line energies, resulting in higher $v_{\rm out}$, at least up to \ion{Fe}{X}. At higher ionisation states, instead, the Fe line energies increase. However, a certain amount of ionisation is required to have non-negligible fractions of non-neutral iron. This would lead to a distribution of ionic abundances and, then, to several emission lines with comparable strength, which are not detected in the present observation. Therefore, our assumption is the most conservative.

To assess the significance of the Fe K$\alpha$ line we simulate $10^4$ XRISM/Resolve spectra using the best-fit model removing the line as input, with same exposure and luminosity as the observation. After fitting the continuum, we search for (spurious) emission features first by fixing the line energy to the observed one and leaving the normalisation and broadening free to vary and, then, leaving also the energy free within a range encompassing the BEL velocities, $0 \leq v_{out} \leq 10^4 \rm km s^{-1}$ (see below), i.e. between 5.40 and 5.59 keV. In the first case, we detect a line with a $\Delta C_{stat}$ improvement higher than our observed line (=11.23) in 4 spectra, implying a statistical significance $>$ 99.9\% (corresponding to $>3 \sigma$), while in the second case in 108 spectra, corresponding to $>98.9$\%.

\begin{table}[]
\centering
\begin{tabular}{l|r }
\hline \hline\\[-5pt]
Parameter & Value \\[1pt]
\hline\\[-5pt]
\textbf{\texttt{galabs}} \\[1pt]
$N_{\rm H}$ (cm$^{-2}$) & $2.0\times 10^{21}$ ($^f$)\\[1pt]
\hline\\[-5pt]
\textbf{\texttt{powerlaw}} \\[1pt]
$\Gamma$ & $2.19 \pm 0.07$ \\[1pt]
norm. $^{a}$ & $2.40\pm 0.02$ \\[1pt]

\hline \\[-5pt]
\textbf{\texttt{gauss$_{\rm wind}$}} \\[1pt]
E (keV) & $5.5^{+1.1}_{-0.2}$ \\[1pt]
$\sigma$ (keV) & $1.8^{+1.0}_{-0.3}$\\[1pt]
norm. $^{b}$ & $1.7^{+1.0}_{-0.2} \times 10^{-4}$ \\[1pt]
\hline \\[-5pt]

\textbf{\texttt{zbfeklor}} \\[1pt]
norm. $^{b}$ & $1.2_{-0.4}^{+0.5} \times 10^{-6}$ \\[1pt]
$\sigma$ (km s$^{-1}$) & $< 310$ \\[1pt]
$z$ & $0.1745_{-0.0008}^{+0.0003}$ \\[1pt]
\hline \\[-5pt]
C-stat/ deg. of freedom & 601.6/592 \\[1pt] \hline
\end{tabular}
\caption{Best-fit values of the 3-6 keV fit of the COR>4 spectrum. ($^f$) denotes a fixed value. Normalisations are in units of: $^a\,10^{-3}\rm\, ph\,keV^{-1}\,cm^{-2}\,s^{-1}$ at 1 keV; $^{b}$ total  $\rm ph\,cm^{-2}\,s^{-1}$ in the line.}
\label{tab:fit}
\vspace{-3ex}
\end{table}

\section{Discussion and Conclusions} \label{sec:discussion}
\subsection{The Fe\,K$\alpha$ line and the Broad Emission Lines}
The BELs of PDS\,456 are known for their extreme velocity and broadening. In the UV, a broad and blueshifted \ion{C}{iv} emission line has been detected in a Hubble Space Telescope/STIS observation by \cite{2005MNRAS.360L..25O} with $v_{\rm out}=5240$ km s$^{-1}$, FWHM$\approx$ 15000 km s$^{-1}$. They also detected unresolved \ion{Si}{iv}/\ion{O}{iv}$\lambda 1400$ lines with $v_{\rm out}=4000$ km s$^{-1}$, FWHM$\approx$ 7000 km s$^{-1}$ and Ly$\alpha$/\ion{N}{v}$\lambda$1240 with $v_{\rm out}=600 - 2000$ km s$^{-1}$ and FWHM$\approx$ 12000 km s$^{-1}$. In the same spectrum, \cite{2018MNRAS.476..943H} claim the presence of a \ion{C}{iv} Broad Absorption Line with $v_{\rm out}=0.3 c$, $\rm FWHM=8550\,km\,s{-1}$, possibly representing a lower-ionisation phase of the powerful X-ray winds.
In the optical band, \cite{1999MNRAS.303L..23S} report a He I line with FWHM$\approx$7000 km s$^{-1}$ and H$\alpha$, H$\beta$ and H$\gamma$ lines with FWHM$\approx$3500 km s$^{-1}$ from a spatially unresolved observation. Similarly, the nuclear region of the MUSE IFU observation in \cite{2024A&A...686A.250T} shows a H$\alpha$ line with FWHM$\approx$2500 km s$^{-1}$ and $v_{out} \approx 350$ km s$^{-1}$. Finally, high-resolution NIR interferometry with GRAVITY resolved the BLR through the Pa$\alpha$ line \citep{2017A&A...602A..94G,2024A&A...690A..76G,2024A&A...684A.167G}. They modelled the BLR as an ensemble of non-interacting "cloudlets" in rough axisymmetric configuration around the central black hole. Such model best-fitted the data through a combination of rotational and outflowing motion. The BLR is located at $r=1.33$\,pc, with an inclination $i=13^\circ$ (almost face-on) with respect to the LOS. Around 50\% of the cloudlets have non-bound orbits, reaching LOS velocities $v_{out} \approx \rm 2000\,km\,s^{-1}$. The total line FWHM is of the same order.

The upper limit for the Fe\,K$\alpha$ FWHM, $1634 (3758)\, \rm km\,s^{-1}$ at 1$\sigma$ (90\%) c.l. is on the low-end side of the above range of values. This may point to a stratified BLR, with closer, broader UV BELs, intermediate optical lines and Fe K$\alpha$ as the outer layer. The Fe K$\alpha$ FWHM can be converted into a rotational radius $r=\frac{GM_{\rm BH}}{(FWHM/\sin(i))^2} \geq 0.014\,\rm pc$, where we adopt the same black hole mass $M_{\rm BH}=1.7 \times 10^8\,\rm M_{\rm \odot}$ and inclination $i=13^o$ (i.e. almost face-on) of GRAVITY for consistency \citep{2024A&A...684A.167G}\footnote{We note that \cite{nrg15} report a quite different $M_{\rm BH}=1.2 \times 10^9\, \rm M_{\rm \odot}$, based on single-epoch BLR scaling relations. Adopting such value would imply a larger rotational radius.}. This value is remarkably similar to the radial location of the X-ray disc wind, between 0.005 and 0.015 pc, estimated in X25, and may indicate that such wind is co-spatial with the BELs, possibly representing a different component which is not gravitationally bound.
To get an estimate on the accretion disc size, we compute the dust sublimation radius $r_{\rm sub}$ as the inner boundary of the surrounding cold "torus". It can be computed as $r_{\rm sub} \approx A_{\rm sub} \sqrt{L_{45}}=4\,\rm pc$, where $L_{\rm bol}/10^{45}\,\rm erg\ s^{-1}$ and $A_{\rm sub}=0.4$ for a sublimation temperature of 1500\,K \citep{1987ApJ...320..537B,2008ApJ...685..160N}.

\subsection{Probing the torus and the X-Ray Baldwin effect at the high luminosity end}
The strength, energy and profile of the Fe K$\alpha$ line is a powerful probe of the covering factor and the column density of the cold reflector. In the Thomson-thin regime, the line flux $I_{Fe\ K\alpha}$ can be related to these quantities as in Eq. 3 of \cite{2001ApJ...546..759Y}:
\begin{equation}
\begin{split}
I_{Fe\ K\alpha}& = 5.6 \cdot 10^{-7} f_{cov}\ N_{\rm H, 22} \Big( \frac{3.2}{\Gamma +1.646} \Big) (7.11)^{1.5 - \Gamma} N_{pl, -3} \\ 
\Rightarrow & f_{\rm cov} \cdot N_{\rm H, 22} =3.7
\end{split}
\end{equation}
where $I_{Fe\ K\alpha}$ is in $\rm ph\ \rm cm^{-2}\ \rm s^{-1}$ (=$1.2 \cdot 10^{-6}$, see above), $N_{\rm H, 22} = \frac{N_{\rm H}}{10^{22}\, \rm cm^{-2}}$, the powerlaw normalisation $N_{pl}$ is in $10^{-3}\rm\, ph\,keV^{-1}\,cm^{-2}\,s^{-1}$ at 1 keV, $f_{\rm cov}=\Delta \Omega / 4 \pi$ is the fraction of the solid angle covered by the reflector. We set $\Gamma=2.3, N_{pl,-3}=3.4$ from the detailed, self-consistent broadband fit of X25. We correct for the updated cross sections as in \cite{2009MNRAS.397.1549M}. Therefore, $N_{\rm H}$ must be of the order of $10^{22} \rm cm^{-2}$ for plausible values of $f_{\rm cov}$; using the GRAVITY best-fit as an example ($f_{\rm cov}=0.67$) we get $N_{\rm H}= 5.5 \times 10^{22} \rm cm^{-2}$. These values are somewhat lower than the "typical" values for BLR $\approx 10^{23} \rm cm^{-2}$ (see e.g. \citealp{of06,netzer13}), in agreement with the Fe K$\alpha$ line tracing the outer layer of a stratified BLR.

In the broad AGN population, the EW of the neutral Fe K$\alpha$ line is found to be anti-correlated with the X-ray continuum luminosity, an effect generally known as the Iwasawa-Taniguchi (or X-Ray Baldwin) effect \citep[e.g.][]{1993ApJ...413L..15I,2007A&A...467L..19B, 2010ApJS..187..581S,2012ApJ...744L..21S}. Although the physical driver is still debated, such effect testifies the evolution of the reflecting matter as a function of the AGN luminosity. 
Our derived EW=9\,eV is significantly smaller than the expected value of 41$\pm$5\,eV obtained through the relation in \cite{2007A&A...467L..19B} (see also \citealp{2010ApJS..187..581S,2012ApJ...744L..21S} for consistent results) for the luminosity of the present observation, $L_{\rm 2-10 keV}=4.7 \times \rm 10^{44}\,\rm erg\ s^{-1}$. PDS\,456 is known to exhibit dramatic spectral and flux variations on timescales as short as 30\,ksec \citep{2002MNRAS.336L..56R}. Albeit quite variable, with a historical average $L_{\rm 2-10 keV} \approx 10^{45}\rm\ erg\ s^{-1}$ \citep{nrg15}, the source sits in the high-luminosity end of the distribution in \cite{2007A&A...467L..19B} \footnote{Formally, luminosities in the \cite{2010ApJS..187..581S,2012ApJ...744L..21S} sample reach $10^{47}\,\rm erg\ s^{-1}$. However, they are computed by rigidly extrapolating a simple, local power law fit in the $2-7$\,keV band and are therefore less accurate than the self-consistent analysis in \cite{2007A&A...467L..19B}.}. Therefore, it is possible that the observed EW may signal a deviation from the established relation at the highest luminosities. We also note that our value is among the lowest measured EWs in the literature, thanks to the unprecedented resolving power of Resolve in the hard X-ray band.

\begin{acknowledgements}
AL, FN acknowledge financial support from grants: EU HORIZON-2020 grant “AHEAD2020” (Agreement No. 871158), ASI Contract No. 2019-27-HH.0 on Athena, PRIN MUR 2022 (DRAGON; No. 2022K9N5B4) and INAF-AF-2023 "The XRISM-to-XIFU (X2X)”, ob.f. 1.05.23.01.06.
\end{acknowledgements}

\bibliography{bibliography.bib}
\bibliographystyle{aasjournal}

\appendix
\section{Comparison between COR>4 and >8 datasets}
\label{appendix}

\begin{figure}[!ht]
\centering
\includegraphics[width=\columnwidth]{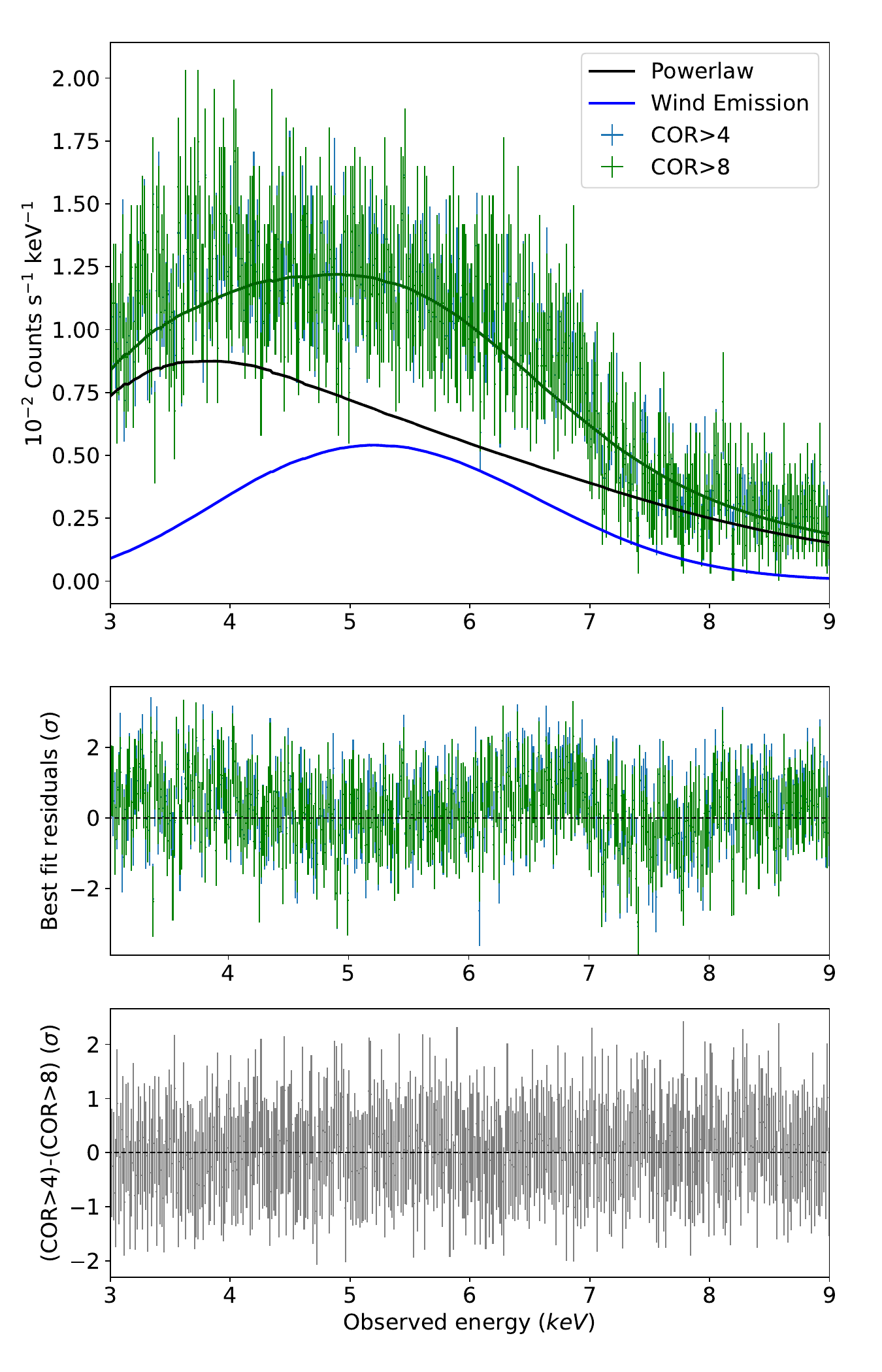}
\caption{XRISM/Resolve spectra and best-fit models (top) and residuals (middle) of PDS\,456 with cutoff rigidity COR>4 and COR>8 (blue and green points, respectively). The best-fit model components are plotted both together and separately, as detailed in the legend. Bottom panels shows the "difference spectrum" between the two datasets. The non X-ray background is not shown here for the sake of clarity.} 
\label{fig:cortime}
\end{figure}
Fig. \ref{fig:cortime} compares the spectra for COR$>$4 and $>$8. Top and middle panels report the data and the best fits for the model in Eq. \ref{cont_model} and the associated residuals. See Sect. \ref{sec:spectra} for details. All best fit values are fully consistent in the two cases. Bottom panel shows the "difference spectrum", computed as the difference between the two spectra, with errors in units of $\sigma$. Such spectrum is consistent with zero, showing that there are no significant differences between the two datasets. The spectra have been binned to 15 eV for visual clarity.

\end{document}